\documentclass[twocolumn,prb,showpacs]{revtex4}

\usepackage{graphics}
\begin{document}
                                                                                
\title{\bf Influence of Domain Wall on Magnetocaloric Effect in
GdPt$_{2}$}

\author{Tapas Samanta and I. Das}

\affiliation{ Saha Institute of Nuclear Physics,1/AF,Bidhannangar,
Kolkata 700 064}
\begin{abstract}
The resistivity, magnetoresistance and in-field heat capacity 
measurements were performed on GdPt$_{2}$ intermetallic compound. 
The magnetocaloric 
parameters $\Delta T_{ad}$ and $-\Delta S$ were derived from the in-field 
heat capacity data. Comparison has been made between the magnetocaloric 
effect $-\Delta S$ and difference in resistivity $-\Delta \rho$
$(=\rho(H)-\rho(0))$ as a function of temperature. There is 
distinct difference in the temperature dependence of $-\Delta S$ 
and $-\Delta \rho$ below the ferromagnetic transition temperature. 
However after removing the domain wall contribution from 
$-\Delta \rho$, the nature of $-\Delta S$ and $-\Delta \rho$ 
dependence as a function of temperature are similar. Our observation 
indicates that the domain wall contribution in magnetocaloric effect 
is negligible in spite of the fact that it has significant contribution 
in magnetotransport.
\end{abstract}
\maketitle
\section{Introduction}
\noindent
The magnetocaloric effect (MCE) is defined as the adiabatic temperature
 change ($\Delta T_{ad}$) or isothermal entropy change ($-\Delta S$) of 
magnetic materials with the application of an external magnetic field. MCE 
has immense technological importance for magnetic cooling. In recent 
years the studies related with MCE has gained momentum due to the 
observation of giant MCE near room temperature \cite{Pecharsky,Fujita,Wada}. 
The main focus in the study of MCE is concentrated to find out newer 
materials with large MCE.
Apart from its technological importance, the MCE can give us valuable 
basic information about the magnetic materials like nature 
of magnetic ordering, metamagnetic transitions etc \cite{Rawat}. 

The building block of ferromagnetic materials below ordering temperature 
are the magnetic domains which are separated by domain walls. MCE is related 
with the thermomagnetic properties of magnetic materials. Therefore magnetic 
domains as well as domain walls is expected to have effect on MCE. However 
the contribution of domain wall on MCE is not properly highlighted 
in the literature.
Polycrystalline GdPt$_{2}$ compound crystallizes in a stable cubic MgCu$_{2}$ 
structure with ferromagnetic Curie temperature 31 K \cite{Modder}. 
In this present work, the main objective is to find out 
how strong is the contribution of domain wall on MCE in GdPt$_{2}$. 
Can it influence the temperature dependence of MCE so much that it leaves 
some strong signature in the dependence.
  
Gadolinium, having $L=0$, has negligible crystalline electric field 
in GdPt$_{2}$ 
and should reach it's full moment value upon ordering and 
attain its full magnetic entropy value $R\ln (2J+1)$ or 17.3 J/mol K. 
Due to large moment of Gadolinium, GdPt$_{2}$ is expected to 
show reasonably large MCE. The magnetic and transport properties of 
GdPt$_{2}$ compound have been studied by various authors 
\cite{Taylor,Kawatra}. It is believed 
that the magnetic interaction of well-localized 4f magnetic moment 
of Gd are mediated by conduction electrons via RKKY interaction. 
Critical behavior of electrical resistivity was studied in the vicinity 
of the ordering temperature in the framework of the molecular field 
theory \cite{Kawatra}. 
To the best of our knowledge no report on the study of thermodynamic property 
of GdPt$_{2}$ compound is available in the literature.  
We have studied MCE as well as magnetotransport properties of GdPt$_{2}$. 
Earlier reports \cite{Das,Das1} in the literature suggest that the dependence
of magnetocaloric effect and magnetoresistance can be similar. The comparison 
of the thermodynamic and magnetotransport data is a novel method of gaining 
deeper understanding about magnetic materials. 
Keeping this context in mind, we have measured and compared the temperature 
dependence of different quantities $-\Delta \rho$ and
$-\Delta S$, one related with transport and the other related
with thermodynamic properties.     

\section{Experimental details}
The binary polycrystalline sample was prepared by arc melting of 
constituent elements of purity better than 99.9$\%$ in Argon atmosphere. 
X-ray diffraction pattern confirms the single-phase nature of the 
compound which crystallizes in cubic MgCu$_{2}$ structure. Specific heat (C) 
measurements were performed using the semi-adiabatic heat-pulse method 
in the temperature interval 4-60 K in the 
presence of 10 and 70 kOe magnetic fields. 
The temperature interval of zero-field 
C measurement was 4-130 K. 
The temperature interval of C measurement in 5 kOe was 4-40 K.  
The temperature dependence of resistivity($\rho$) in the 
absence of a field as well as in the presence of 5, 10 and 70 kOe magnetic 
fields were measured by the conventional four-probe method. 
The longitudinal magnetoresistance (MR)
$\left(\Delta\rho/\rho=\{\rho(H)-\rho(0)\}/\rho(0)\right)$ 
measurements at 4, 10, 20, 40 and 80 K were carried out in the 
magnetic field up to 75 kOe.
\section{Results and discussion}
The specific heat of GdPt$_{2}$ as a function of temperature at various 
constant magnetic field is plotted in Fig.1.  
\begin{figure}[ht]
\resizebox{8.5cm}{7.5cm}
{\includegraphics{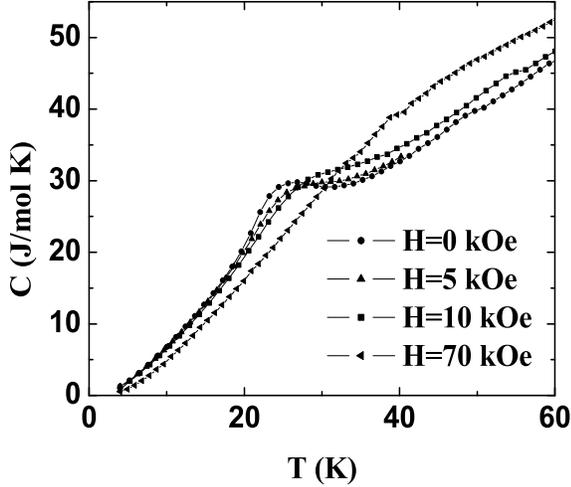}}
\caption{Heat capacity (C) as a function of temperature for GdPt$_{2}$ at 
different constant magnetic fields}
\end{figure}
In the presence of small external
magnetic field ($\simeq$10 kOe) the peak position of C shifts to the higher
temperature, indicating the ferromagnetic nature of magnetic ordering.
At higher magnetic field i.e. in 70 kOe field, the peak disappears 
completely.

To find out the magnetic contribution of specific heat 
we have fitted the zero-field C 
data using Debye integral along with linear contribution within the 
temperature interval 80 to 130 K and extrapolated the fitted data down to low
temperature which is shown in Fig.2. The total specific heat C can be 
expressed as,
\[C=C_{el}+C_{ph}+C_{mag}\]
C$_{mag}$ is the magnetic part of specific heat. C$_{el}$ and C$_{ph}$ 
are respectively the electronic and phonon contribution of 
specific heat. The electronic part is of the form $C_{el}=\gamma T$ 
,where $\gamma$ is the electronic heat capacity coefficient. 
The phonon part, approximated as Debye model, is of 
the form $C_{ph}=\mathcal{D}(\theta_{D}/T)$, 
where $\mathcal{D}(\theta_{D}/T)$ is the Debye function and 
$\theta_{D}$ is the Debye temperature.
\begin{figure}[ht]
\resizebox{8.5cm}{7.5cm}
{\includegraphics{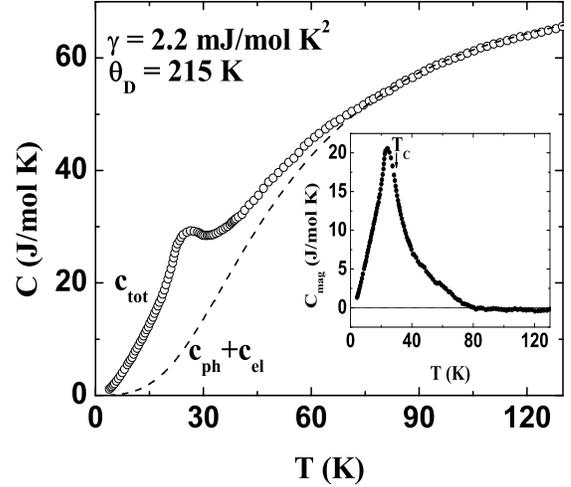}}
\caption{Zero-field specific heat data as a function of temperature. 
Dashed line represents the lattice contribution of specific heat.  
Inset: magnetic contribution of specific heat as a function of temperature.}
\end{figure}
The C data was fitted using
\[C_{el}+C_{ph}=\gamma T+\mathcal{D}(\theta_{D}/T)\]
in the temperature interval 80 to 130 K under the approximation that well above 
the transition temperature magnetic contribution is negligibly small. 
From fitting, the value of $\gamma$ and $\theta_{D}$ turns out to 
be 2.2 mJ/mol K$^{2}$ and 215 K respectively. Magnetic contribution of 
specific heat was obtained by subtracting the regenerated nonmagnetic 
contribution in the temperature range 4 to 130 K using the above 
mentioned $\gamma$ and $\theta_{D}$ value.
The temperature dependence of $C_{mag}$ is shown in the inset of Fig.2. 
From the inflection point of $C_{mag}$ data, we obtained the ferromagnetic 
ordering temperature $T_{C}=29$ K which is close to the referred transition 
temperature $T_{C}=31$ K \cite{Modder} from magnetization measurement. The 
maximum value of $C_{mag}$ reaches 20.54 J/mol K. The magnetic contribution 
to the specific heat for equal-moment (EM) magnetic structure in Gd 
intermetallic compounds is expressed as \cite{Blanco}
\[C_{EM}=\frac{5J(J+1)}{(2J^{2}+2J+1)}R\]
Gadolinium having J=7/2, yield $C_{EM}=20.15$ J/mol K. 
Our experimentally observed value of $C_{mag}$ which is very close to 
the $C_{EM}$ value indicates that the magnetic configuration in GdPt$_{2}$ is 
equal-moment in nature. Moreover a noticeable magnetic contribution 
persists well above the transition temperature.
The magnetic entropy of Gd intermetallic compounds attain its full 
value $R\ln(2J+1)$ or 17.3 J/mol K just above the ordering temperature 
 \cite{Bouvier}. The calculated magnetic entropy of our sample is 
17.6 J/mol K at the ordering temperature which is in good 
agreement with $R\ln8$ or 17.3 J/mol K. This indicate that the Gd ions 
ordered with full moments within GdPt$_{2}$.

The isothermal entropy change ($-\Delta S$) and adiabatic 
temperature change ($\Delta T_{ad}$) was obtained from total entropy, 
which was calculated from experimental C data as a function of 
temperature at various constant magnetic fields. 
To calculate the entropy contribution for 0 to 4 K, the linear 
variation of C data was considered. The difference between the two 
entropy curves from zero-field to in field for isothermal translation results 
in $-\Delta S$ and isentropic subtraction gives $\Delta T_{ad}$. 
The temperature dependence of $-\Delta S$ and $\Delta T_{ad}$ for 5, 10 
and 70 kOe magnetic fields are plotted in Fig.3 and Fig.4 respectively.
\begin{figure}[ht]
\resizebox{8.5cm}{7.5cm}
{\includegraphics{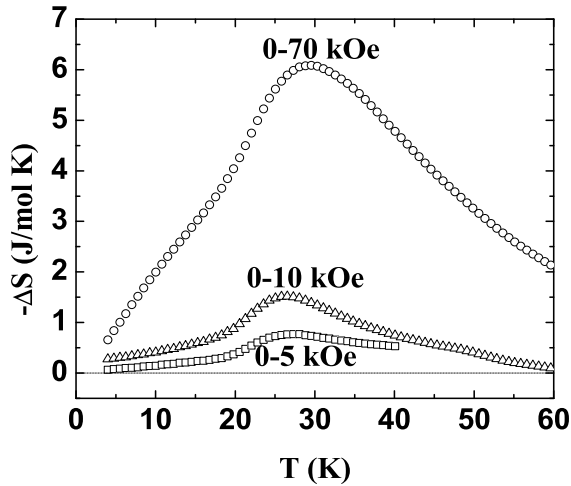}}
\caption{Isothermal entropy change $-\Delta S$ as a function of temperature
calculated from the heat capacity data at constant magnetic fields.}
\end{figure}
\begin{figure}[ht]
\resizebox{8.5cm}{7.5cm}
{\includegraphics{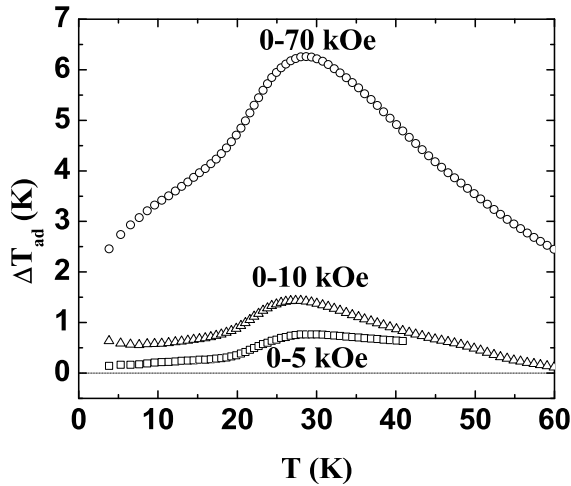}}
\caption{Adiabatic temperature change  $\Delta T_{ad}$ as a function of 
temperature calculated from the heat capacity data at constant magnetic fields.}
\end{figure}
The plot of $\Delta T_{ad}$ as a function of temperature shows a positive 
caret-like shape with maxima around the magnetic ordering temperature and 
$\Delta T_{ad}$ positive in the entire temperature range for all magnetic 
fields, which is expected for ferromagnetic materials. 
The temperature dependence of both 
$\Delta T_{ad}$ and $-\Delta S$ are almost similar to each other. 
The value of $\Delta T_{ad}$ around the magnetic ordering temperature 
for 5, 10 and 70 kOe magnetic fields are respectively 
0.8, 1.4 and 6.3 K i.e. the rate of change of $\Delta T_{ad}$ as a function 
of magnetic field decreases with increasing fields. This feature also 
indicate the ferromagnetic nature of GdPt$_{2}$ compounds.

\begin{figure}[ht]
\resizebox{8.5cm}{7.5cm}
{\includegraphics{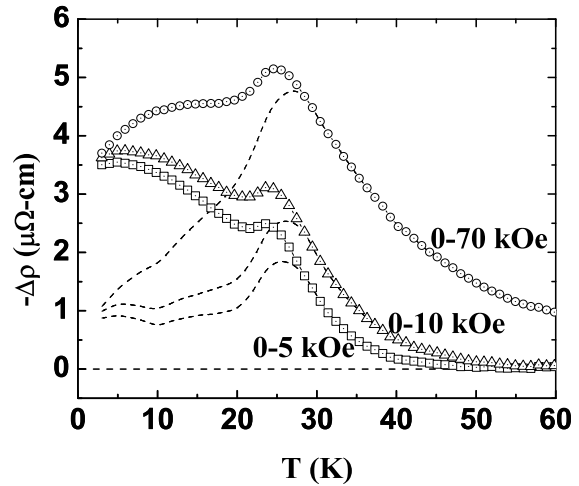}}
\caption{Differences in resistivity $-\Delta \rho$ are plotted as a function 
of temperature with different symbols. Dashed line curves are the 
$-\Delta \rho$ vs. temperature curve after subtracting domain wall 
contribution in magnetoresistance.}
\end{figure}
The temperature dependence of $-\Delta \rho$ is shown in Fig.5 which
was calculated from experimental resistivity data from 4 to 60 K at
various constant magnetic fields. Below the magnetic ordering temperature the
variation of $-\Delta \rho$ and $\Delta T_{ad}$ or $-\Delta S$
with temperature are distinctly different for all the three magnetic fields
5, 10 and 70 kOe.
It has been shown earlier that the temperature dependence of 
$-\Delta \rho$ and $-\Delta S$ can be similar \cite{Das,Das1}. 
It implies that for a ferromagnetic compound 
with the increasing (decreasing) magnitude of $-\Delta S$ the 
magnitude of $-\Delta \rho$ is expected to increase (decrease) as a 
function of temperature. As a result one can expect that $-\Delta \rho$ 
decreases gradually as does $-\Delta S$ with decreasing temperature after 
showing a maxima around ferromagnetic transition temperature of GdPt$_{2}$. 
In contrast to the expectation, $-\Delta \rho$ shows a broad hump at low 
temperature. To find out the main cause behind the dissimilar behavior between
$-\Delta \rho$ and MCE we have performed MR measurement as function of field 
at different constant temperatures, which is shown in Fig.6(A).
\begin{figure}[ht]
\resizebox{8cm}{9cm}
{\includegraphics{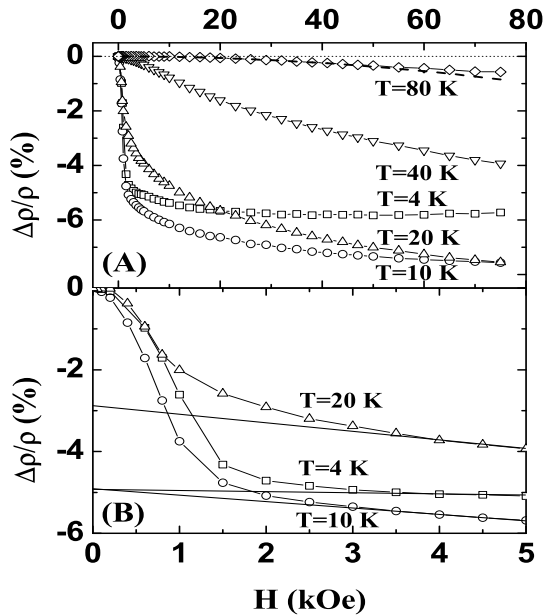}}
\caption{(A) Magnetoresistance as a function field at various constant 
temperatures. (B) LFMR at different temperature calculated by 
extrapolating 5 kOe data to zero field.}
\end{figure}
The MR curves at constant temperature clearly demonstrate the existence 
of significant low field magnetoresistance (LFMR) originating from 
magnetic domain wall at low temperature. In the 
paramagnetic state the low field MR vanishes and at higher temperature 
i.e. at 80 K MR follows -$H^2$ magnetic field dependence
as indicated by dashed line in Fig.6(A), which is an indication of enhanced 
spin fluctuation even at this high temperature. 
The LFMR value was obtained by extrapolating 5 kOe data to zero field
which is shown in Fig.6(B). The LFMR below ferromagnetic ordering 
temperature of polycrystalline compounds 
originates due to the suppression of domain wall scattering of conduction 
electrons with the application of magnetic field. 
Domain wall contribution of MR in GdPt$_{2}$ remains unchanged below 10 K and 
absent above the ferromagnetic transition temperature. In the intermediate 
temperatures it varies almost linearly as a function of temperature. 
After removing the domain wall contribution from the total resistivity 
difference the broad 
hump in $-\Delta \rho$ vanishes which is shown in Fig.5 by dashed line
and the nature of $-\Delta \rho$ 
and $-\Delta S$ curves as a function of temperature comes out to be similar. 
These observation indicates that the dissimilar temperature dependence 
of $-\Delta \rho$ and $-\Delta S$ in GdPt$_{2}$ is originating from 
the fact that the magnetic domain wall has significant contribution 
in $\Delta \rho$ but negligible influence on MCE.
 
\section{Summary}
MCE along with transport property have been studied in GdPt$_{2}$ 
compound. We have observed distinct difference in temperature dependence 
of $-\Delta \rho$ and $-\Delta S$ below the ferromagnetic ordering temperature. 
However if we remove the domain wall contribution from 
$-\Delta \rho$, then the nature of $-\Delta \rho$ and $-\Delta S$ 
curves as a function of temperature are similar. It highlights the fact 
that the domain wall contribution in magnetocaloric effect is negligible 
in spite of the fact that it has significant contribution in transport 
in GdPt$_{2}$.


\end{document}